\documentclass[11pt,aps,prd,preprint,preprintnumbers,showpacs,superscriptaddress,floatfix]{revtex4-1}
\usepackage{graphicx}
\usepackage{epsfig}
\usepackage{dcolumn}
\usepackage{bm}
\usepackage{subfigure}
\usepackage{amsmath}
\usepackage{amssymb}
\usepackage{bbm}
\usepackage{multirow}
\usepackage{setspace}
\usepackage{xcolor}
\usepackage[colorlinks,
            linkbordercolor={0 0 0},
            pdfborder={0 0 0},
            linkcolor=blue,
            citecolor=blue,
            urlcolor=blue,
            breaklinks=true]{hyperref}

\allowdisplaybreaks

\newlength{\x}
\newlength{\y}
\newlength{\z}
\phantomsection

\begin{document}

\title{Effect of  fluctuations on the Geodesic rule for topological defect formation}

\author{Sanatan Digal}
\email{digal@imsc.res.in}
\affiliation{The Institute of Mathematical Sciences, Chennai 600113, India}
\affiliation{Homi Bhabha National Institute, Training School Complex,
Anushakti Nagar, Mumbai 400094, India}

\author{Vinod Mamale}
\email{mvinod@imsc.res.in}
\affiliation{The Institute of Mathematical Sciences, Chennai 600113, India}
\affiliation{Homi Bhabha National Institute, Training School Complex,
Anushakti Nagar, Mumbai 400094, India}

\begin{abstract}
\vskip 1cm
At finite temperature, the field along a linear stretch of correlation length size is supposed to trace the shortest path in the
field space given the two end point values, known as the Geodesic rule. In this study, we compute the probability that, the field variations over distances of correlation length follow this rule in theories with $O(2)$ global symmetry. We consider a simple ferromagnetic $O(2)$ spin-model and a complex $\phi^4$ theory. The computations are carried out on an ensemble of equilibrium configurations, generated using Monte Carlo simulations. The numerical results suggest significant deviation to the Geodesic rule, relevant for formation of topological defects during quench
in 2nd order phase transition. Also for the case of $O(2)-$spins in two dimensions, distribution and density of vortices, have been studied. 
It is found that, for quench temperatures close to the transition point, the Kibble-Zurek Mechanism underestimates equilibrium density of defects. 
The exponents corresponding to width of the distributions, are found to be smaller than Kibble Mechanism estimates and match only when there is no deviation from the geodesic rule.

\end{abstract}

\maketitle
\section{Introduction} 
\label{sec:intro}  
 
Topological defects arise in a wide variety of systems, ranging from  table-top experiments in condensed matter systems\cite{Chuang:1991zz, Hendry:1994,Ruutu:1996,Bauerle:1996,Dodd:1998aan,Carmi:2000zz,Digal:1998ak} to theories of the early Universe, 
extremely dense stars etc.\cite{Hindmarsh:1994re,Magueijo:2000se,Srivastava:2017itj}. They are non-zero energy static solutions 
when there is spontaneous symmetry breaking(SSB). The processes of formation and evolution of these defects have been extensively studied in the literature. There are many theoretical as well as experimental studies available on the formation of topological defects in condensed matter systems \cite{Srivastava:2001,Chaikin2000}. In the case of high energy physics the studies are mostly theoretical. However, since topological considerations play a dominant role in their formation and evolution, ideas proposed for high energy physics systems, have been tested in condensed matter systems \cite{Chuang:1991zz, Hendry:1994,Ruutu:1996,Bauerle:1996,Dodd:1998aan,Digal:1998ak,Bowick:1992rz,Ray:2001ug,Rajantie:2001ps}.

\begin{figure}[!tbp]
  \centering
  \begin{minipage}[b]{0.75\textwidth}
  \hskip-1.0cm
    \includegraphics[width=0.75\textwidth]{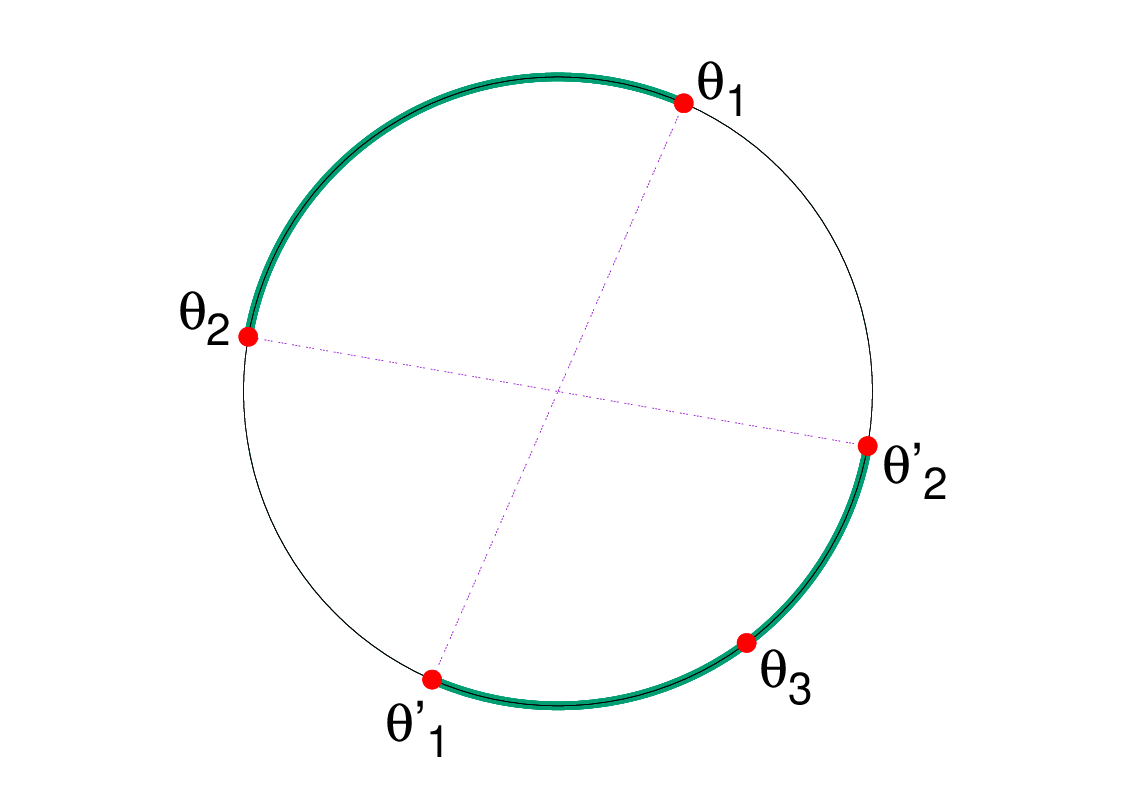}
    \caption{Implementation of geodesic rule in two spatial dimensions, with circle as an OPS.}
 \label{geoo}
 \end{minipage}
\end{figure}

The theory of formation of topological defects was proposed in the context of the early Universe by T. W. B. Kibble, known as the Kibble-Mechanism(KM) \cite{Kibble:1976sj}. The Kibble-Mechanism is based on two postulates. In the immediate aftermath of a phase transition, when
there is SSB, the order parameter (OP) in physical space takes values from the order parameter space (OPS).  According to the first postulate, the physical space splits into ``uncorrelated" domains, inside which the OP is roughly uniform. The second postulate, which is also known as the ``Geodesic Rule", states that in between adjacent domains, OP field interpolates along the shortest path in the OPS \cite{Kibble:1995aa,Kibble:1976sj}. Using
these two postulates the defect densities in terms of the domain size $\xi$ can be calculated. For example, when OPS is a circle and OP is
specified by phase $\theta$, minimum three domains are required to form a vortex or a string segment. The corresponding area is $\sim\sqrt{3}\xi^2/4$. The probability of formation of a defect near the junction of these three domains, is calculated as follows. Suppose the phase of the order parameter in two of the domains is $\theta_1$ and $\theta_2$. $\theta^\prime_1$ and $\theta^\prime_2$ are diametrically opposite to $\theta_1$ and $\theta_2$ on the OPS circle. The geodesic rule restricts the variation of $\theta$ between the domains to $|\theta_2 -\theta_1| \le \pi$. The rule also restricts $\theta_3$ to lie on the geodesic between $\theta^\prime_1$ and $\theta^\prime_2$ in order that there is a defect, as
shown in Fig.\ref{geoo}. So the probability of having a defect is,
\begin{equation}
P = {\int_0^\pi (q/2\pi) dq\over \int_0^\pi dq} = {1 \over 4},
\end{equation}
where $q=|\theta_1-\theta_2|$ which can vary between $0$ and $\pi$. Hence, density of defects, $\rho$, is ${\xi^{-2}/\sqrt{3}}$. Note that, the factor ($1/\sqrt{3}$) will change with geometrical considerations. We mention here that, $\rho\sim \xi^{-2}$, has also been considered on dimensional grounds ~\cite{Kibble:1980mv,Vilenkin:1981kz}.

The KM has been successfully applied to study formation of defects in condensed matter systems as well as in numerical experiments in the context of the early Universe \cite{Chuang:1991zz, Hendry:1994,Ruutu:1996,Bauerle:1996,Dodd:1998aan,Digal:1998ak,Bowick:1992rz,Ray:2001ug,Rajantie:2001ps}. In a first order phase transition, bubbles of the SSB phase nucleate in the background of symmetric phase \cite{Srivastava:1991nv}. These bubbles can be identified with the domains proposed in the Kibble Mechanism. It has been observed in numerical
simulations that the OP between colliding bubbles traces the shortest path in the OPS validating the Geodesic rule \cite{Srivastava:1991nv,Chakravarty:1992zp}. Though there can be exceptions, when kinematics allow energetic bubble collisions or when there is a small explicit symmetry breaking present apart from the SSB \cite{Copeland:1996jz,Digal:1997ip,Digal:1995vd}.  

In a second order phase transition formation of topological defects is described by the Kibble-Zurek mechanism (KZM)\cite{Zurek:1985qw, Zurek:1996sj,Yates:1998kx}. In this case, it becomes necessary to take into account the relaxation time ($\tau$) of the system, which diverges at the critical temperature $T_c$, i.e
\begin{equation}
\tau = {\tau_0 \over |\epsilon|},~~ \epsilon= (T - T_c)/T_c,
\end{equation}
where $\tau_0$ is a constant. In ref.\cite{Zurek:1985qw}, Zurek considered a quench, $\epsilon(t) = t/\tau_Q$, where $\tau_Q$
is the quench time scale. Zurek argued that, the system remains in the adiabatic regime with the OP
keeping up with changes in temperature for, $|t|\ge \tau$.  On the other hand for $|t| < \tau$, the OP effectively freezes out. The freeze out starts,
when $|t_{fr}|=\tau$, i.e at time $t_{fr} = \sqrt{\tau_0\tau_Q}$. At $t_{fr}$, the temperature $T_{fr}$ and the equilibrium
correlation length $\xi_{fr}$ are given by,
\begin{equation}
T_{fr} = T_c + \delta,~~\xi_{fr} \sim \xi_0 \left(\tau_0 \over \tau_Q\right)^{-\nu/2},
\end{equation}
where $\delta = T_c\sqrt{\tau_0/\tau_Q}$, $\xi_0$ is a constant which depends on the system. The critical exponent $\nu$ depends on the universality class . The freeze out continues until the system cools down to temperature $\sim T_c-\delta$, when the relaxation time becomes small enough such that the field can keep up with change in temperature. Thus, the density of defects as the system comes out of the above non-adiabatic regime, is given by~\cite{Zurek:1996sj,Kibble:1980mv,Vilenkin:1981kz}
\begin{equation}
\rho_{fr} \sim {1 \over \xi_{fr}^2}.
\label{dn1}
\end{equation}
This power law, depends on the domain size and the Geodesic rule~\cite{Kibble:1976sj,Kibble:1995aa}. We mention here that, the hard quench considered above is an approximation. As a result, the density of defects when the defects stabilise will differ from Eq.\ref{dn1} ~\cite{Karra:1998gn}. In some studies this is taken care by introducing pre-factors ~\cite{delCampo:2013nla,Dora:2019,Chesler:2014gya}. Considering such a prefactor does not reduce the discrepancy in density of defects discussed in section-II(D). There also have been studies in which an initial equilibrium density of defects is evolved in time, as the system cools, by including thermal fluctuations to give the effect of partial thermalisation~\cite{Asja Jelic}.

In the present work, we study the order parameter variations and density of defects for a system in equilibrium. 
In particular we study, whether the order parameter traces the shortest (geodesic) path for different spatial separations. From this, effects of fluctuations on the Geodesic rule of KM can be estimated. The temperature dependence of the density of defects are compared with conventional estimates from the $KM$. Note that $T_{fr}$ can take any value above $T_c$ depending on the cooling rate of the system. Hence different possible values of $T_{fr} > T_c$ are considered. For convenience, the subscript $``fr"$ will be dropped in the rest of the paper. The study is mostly on $O(2)-$spin model in two and three spatial dimensions. We also consider $\phi^4$ theory with $U(1)\equiv O(2)$ symmetry in three spatial dimensions to see the effect of radial fluctuations. The effect of fluctuations are computed by simulating the partition functions, using conventional Monte Carlo methods. To measure the deviation
at spatial separation $r$, the field configuration over a linear stretch of $r$ on the lattice, is mapped to a path in OPS and it's length is computed.
A geodesic is drawn between the two end points of this path.  When there is a deviation to the Geodesic rule, the two paths mostly lie on the opposite sides of OPS. The length of Geodesic is always $\le \pi$. There is no restrictions on the actual
path length, some paths can even wind the OPS more than once. For, $r=\xi_{KM}$ no preference to the Geodesic path is found. The deviation to the geodesic rule quantitatively depends on the definition of the domain size. However, as long as the domain size is larger than the lattice spacing, there will be non-zero deivation. It is found that even the maximal deviation from the Geodesic rule largely underestimates the defect densities, which suggests contribution of thermal fluctuations dominate the density of defects.

The paper is organised as follows. In section-II and section-III, numerical calculations to determine the validity
of geodesic rule are presented for the $O(2)-$spin model  and $\phi^4$ theory respectively. The conclusions are presented in
section-IV.

\section{Geodesic rule in $O(2)-$spin model}

The $O(2)-$spin model, also known as the $XY-$model, is defined on the lattice \cite{Kosterlitz:1973xp}. The Hamiltonian of the model is 
given by,
\begin{equation}
H = -J \sum_{\left<ij\right>} \vec{s}_i.\vec{s}_j.
\end{equation}
where $\vec{s}_i$ is the spin at the site $i$, ${\left<ij\right>}$ denotes nearest neighbour (NN) spin pairs. The coupling $J$ is ferromagnetic, i.e $J>0$.  The $O(2)-$symmetry corresponds to global rotations of the spins which preserve the Hamiltonian. The partition function is given by,
\begin{equation}
{\cal{Z}} = \int \prod_i^{N} d\vec{s}_i Exp[\beta \sum_{\left<ij\right>} \vec{s}_i.\vec{s}_j].
\label{prt1}
\end{equation}
$\beta = J/\kappa T$, where $\kappa$ and $T$ are the Boltzmann constant and temperature respectively. For convenience $T$ is expressed in units of $J/\kappa$. The observables computed are mangetisation ($\vec{M}$), energy $E$, defined as,
\begin{equation}
\vec{M} = {1 \over V} \sum_i \vec{s}_i,~E=-J \sum_{\left<ij\right>} \vec{s}_i.\vec{s}_j
\end{equation}
where $V$ is total number of spins. The magnetic susceptibility, $\chi$, defined as,
\begin{equation}
{\chi}(T) = \langle |\vec{M}|^2\rangle - \langle |\vec{M}| \rangle ^2
\end{equation}
is used to locate the transition point. It is well known that the above model undergoes a $2$nd order ferromagnetic-paramagnetic transition at critical temperature $T_c$. The vortices are present even above $T_c$ though they are unstable. Below $T_c$, they would be stable but can not be observed in this equilibrium set up. In the case of two physical dimensions, there is  no magnetisation transition due to the Goldstone modes. Interestingly, the system undergoes the well known Kosterliz-Thouless(KT) transition\cite{Kosterlitz:1973xp}. The spin system consists of a network of vortices at all temperatures. The high temperature phase is described by a screened Coulomb gas of vortices and a strongly interacting vortex pairs below the KT transition. Though the two dimensional system is not directly relevant for the formation of topological defects via the Kibble-Zurek mechanism, the distribution of vortices at high temperature serves as an instructive example. There is no reason to expect that the Geodesic rule will not hold in the case of two spatial dimensions. We mention here that, defects in the above model in two dimensions have been studied in the context of KZM~\cite{Asja Jelic}. The defect distributions are evolved in time, taking into account decrease in the temperature. Effects of transient temperatures are taken into account by  considering Monte Carlo evolution of the system.

The thermodynamic observables corresponding to the partition function, Eq.\ref{prt1}, are computed using numerical Monte Carlo simulations.
In simulations, a sequence of statistically ``independent" configurations, Monte Carlo history, is generated using the standard metropolis algorithm\cite{Hasting:1970}. In this algorithm, an initial configuration of spins is randomly selected. Thereafter, each spin is updated sequentially. Single sweep is done when all spins are considered for updating. In the Monte Carlo simulations, since a new  configuration of spins is generated from old one, the two are correlated. To reduce this auto-correlation, $\sim 10$ sweeps are carried out before accepting a configuration. The resulting sequence of configurations constitute the Monte Carlo history. The partition function average of physical observables are then given by average over the Monte Carlo history.
For computations we consider about $10^5$ configurations for each temperature. The error analysis is carried out using standard Jacknife method.
For simplicity, a square lattice ($N = N_s\times N_s$) and a cubic lattice ($N=N_s\times N_s \times N_s$) are considered for two and three dimensions respectively. 

\subsection{Calculation of $\xi$ and $\xi_{KM}$}

Conventionally the correlation length $\xi$ is obtained from the correlation function, 
\begin{equation}
c(r)=\left\langle \vec{s}_i\cdot \vec{s}_j\right\rangle - \left\langle \vec{s}_i \right\rangle \cdot \langle\vec{s}_j\rangle,~~r=|{\rm \bf r}_i-{\rm \bf r}_j|.
\end{equation}
For large $r$, $c(r)$ is expected to decay as $\sim Exp[-r/\xi(T)]$. Clearly at $r=\xi$, $c(r)\sim 1/e$, consequently $\vec{s}_i$ can not take values independent of $\vec{s}_j$ and vice-versa. According to KM the partition average of $\Delta \theta = cos^{-1}(\vec{s}_i\cdot \vec{s}_j)=\rm{min}\left[|\theta_i-\theta_j|, 2\pi-|\theta_i-\theta_j|\right]$ and  $\langle \theta_i\theta_j\rangle$ should be $\pi/2$ and $\pi^2$ respectively. 
At $T=1.18$, $\xi \simeq 4.4$ in lattice units. At these separations, it is found that $\langle\Delta \theta\rangle\approx 0.87\pi/2$ and $\langle \theta_i\theta_j\rangle\approx 1.03\pi^2$. We also consider $\xi_{KM}$ as the least physical separation for which $\langle\Delta \theta\rangle\simeq \pi/2$.  At $T=1.18$, it so happens that $\xi_{KM}$ is larger than $\xi$ by a factor of $\sim 5$.

\subsection{The Geodesic rule over $\xi$ and $\xi_{KM}$ separations}

The calculations to check the validity or the deviations of the Geodesic rule necessitates that image trajectory of field configurations in the field
space are defined. The image in the field space will just be a set of points, since the field is defined only on the lattice grid. To draw a continuous trajectory through these points, the Geodesic rule between NN points is assumed. In the field space, a pair of points, which correspond to NN sites
on the lattice, are connected by
a line/arc. We mention here that the Geodesic rule is taken into account while discretising a continuum theory on the lattice. Also in numerical simulations, this rule is used to locate as well as compute the windings of the defects.

In the present case the field space or the OPS is a circle, which can be characterised by an angular variable $\theta \in [0-2\pi]$ with condition
$\theta=\theta+2\pi$. If the field configuration between $\vec{s}_i$ and $\vec{s}_j$ follows the Geodesic rule then the length of the image trajectory will correspond to $\Delta\theta_{ij}=\rm{min}[|\theta_i-\theta_j|,2\pi-|\theta_i-\theta_j|]$. The actual length corresponding the field configuration from ${\rm \bf r}_i$ to ${\rm \bf r}_j$, the sites of $\vec{s}_i$ and $\vec{s}_j$ respectively, is given by the absolute value of, 
\begin{eqnarray}
\eta_{ij}(r=|{\rm \bf r}_i-{\rm \bf r}_j|)=\sum_{k=i}^{j-1} \alpha_k  \Delta\theta_{k+1,k},~~~~~~~~~~~~~~ \nonumber\\
\Delta\theta_{k+1,k}=\rm{min}\left[|\theta_{k+1}-\theta_k|,2\pi-|\theta_{k+1}-\theta_k|\right].
\end{eqnarray}
$\alpha_k=+1(-1)$, if the trajectory traces a path clockwise(anti-clockwise) in OPS as one goes from site ${\rm \bf r}_k$ to ${\rm \bf r}_{k+1}$.
Note that $\eta_{ij}(r)$ corresponds to net variation of angle $\theta$ of the spins over $r$ separation. We compute the partition average of $d(r)$, the deviation to the Geodesic rule at separation $r$, defined as
\begin{equation}
d(r) = \left\langle {1 + \gamma \over 2 }\right\rangle 
\end{equation}
where $\gamma=1$ if $|\Delta\theta_{ij}(r)|<|\eta_{ij}(r)|$; else $\gamma=-1$. In case $\Delta\theta_{ij}\ne \eta_{ij}$ at a given $r$, it is counted as the deviation from the geodesic rule. This computation is repeated for $r$ up to $\xi_{KM}$.

\begin{figure}[!tbp]
  \centering
  \begin{minipage}[b]{0.485\textwidth}
    \includegraphics[width=\textwidth]{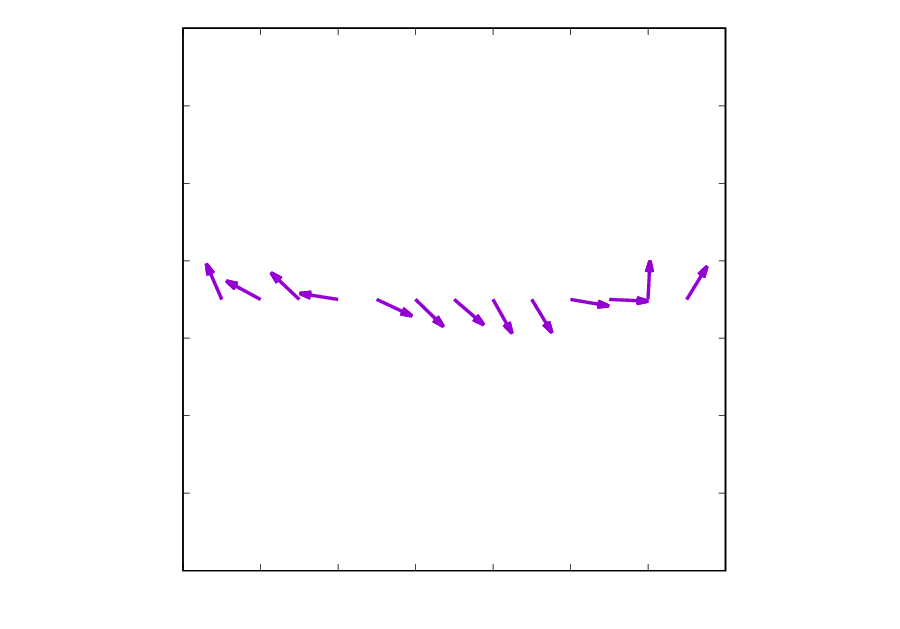}
    \caption{A configuration spins over length $\xi$ in physical space}
 \label{fig3}
  \end{minipage}
   \hspace{0.3cm}
  \begin{minipage}[b]{0.47\textwidth}
    \includegraphics[width=\textwidth]{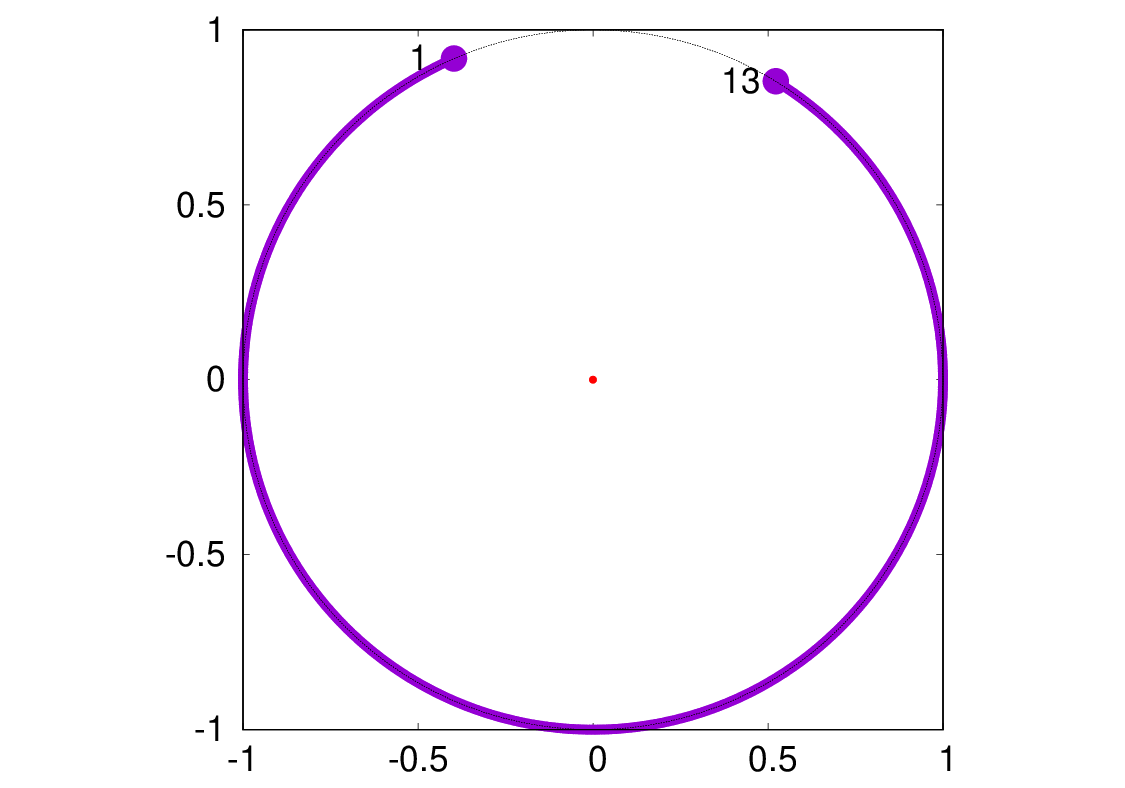}
    \caption{The corresponding trajectory $\eta_{1,13}$ defined in Eq.(5) on
    the OPS}
   \label{fig4}
  \end{minipage}

  \begin{minipage}[b]{0.485\textwidth}
    \includegraphics[width=\textwidth]{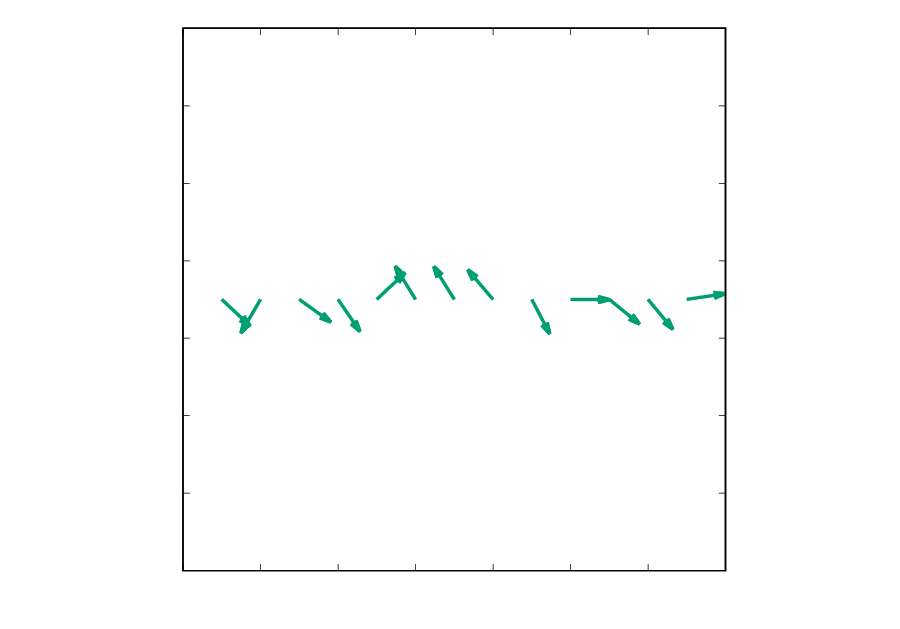}
    \caption{A configuration spins over length $\xi$ in physical space}
 \label{fig5}
  \end{minipage}
   \hspace{0.3cm}
  \begin{minipage}[b]{0.47\textwidth}
    \includegraphics[width=\textwidth]{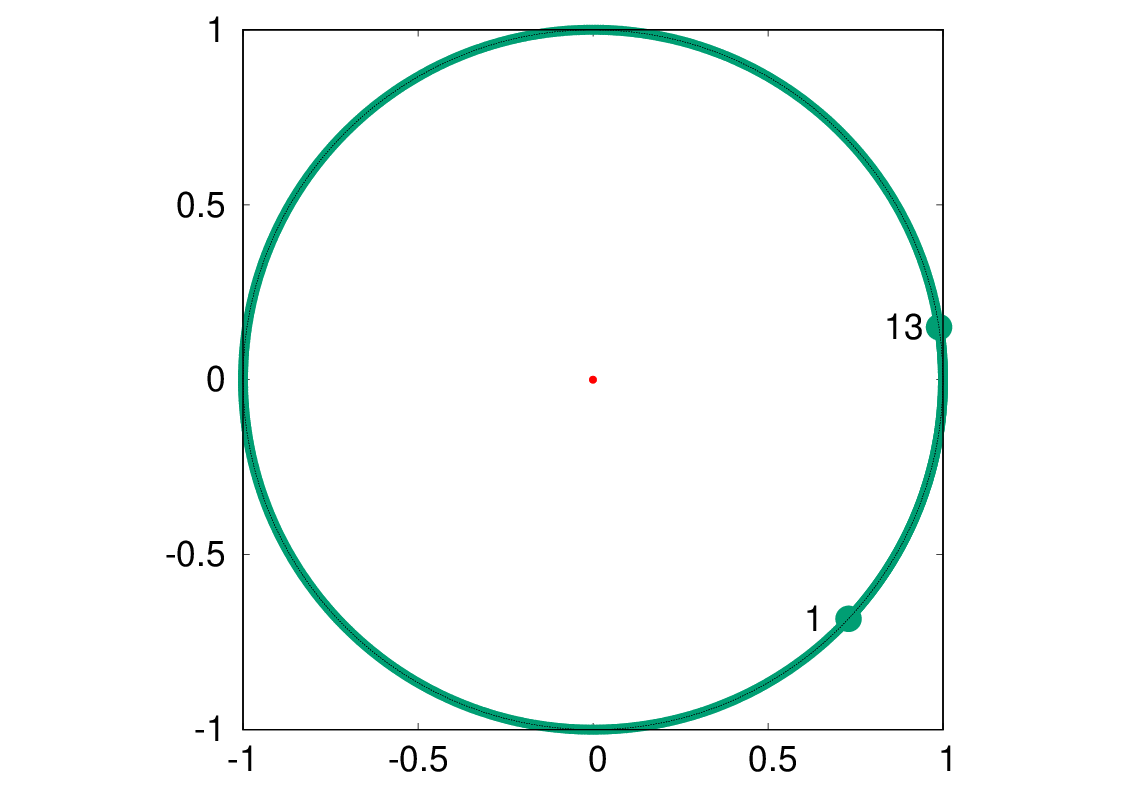}
    \caption{The corresponding trajectory $\eta_{1,13}$ defined in Eq.(5) on
    the OPS}
   \label{fig6}
  \end{minipage}
\end{figure}

In figure \ref{fig3} we show an example of the field configurations along a linear stretch of  $\xi_{KM}$ for $T=1.43$. $\xi_{KM}=12$ for this $T$. The corresponding image in the OPS is shown in figure \ref{fig4}. The two end points correspond to $\theta_1 \simeq 0.63\pi$ and $\theta_{13} \simeq 0.324\pi$, with $\Delta\theta_{1,13}=0.306\pi$. $\eta_{1,13}$ in this case turns
out to be $2\pi-\Delta\theta_{1,13}=1.694\pi$; which does not correspond to the shortest path. This configuration is an example
where deviation from the Geodesic rule is observed. In figures \ref{fig5} another configuration is shown  for the same $T$. The corresponding image map is shown in \ref{fig6}. The endpoints for this configuration are; $\theta_1 \simeq 1.76\pi$ and $\theta_{13} \simeq 0.05\pi$.  In this case $\Delta\theta_{1,13}=.29\pi$ but $\eta_{1,13}$ is found to be $2.29\pi$. This trajectory not only deviates from the Geodesic rule but also winds the field space, i.e. OPS, once. This implies that for given two end points in the field space there are many ways in which the Geodesic path can be avoided, but there is only one way it is followed.

\begin{figure}[!tbp]
  \centering
  \begin{minipage}[b]{0.45\textwidth}
    \includegraphics[width=\textwidth]{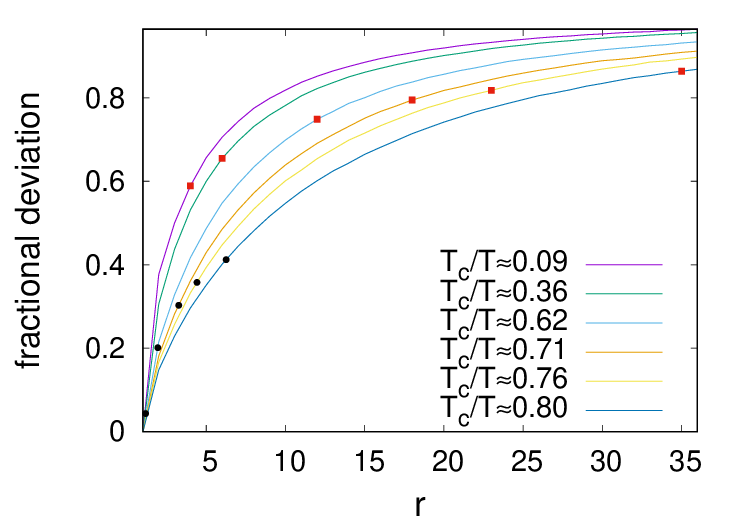}
    \caption{Geodesic rule devation in $2D-$XY model with lattice distance}
 \label{fig7}
  \end{minipage}
   \hspace{0.3cm}
  \begin{minipage}[b]{0.45\textwidth}
    \includegraphics[width=\textwidth]{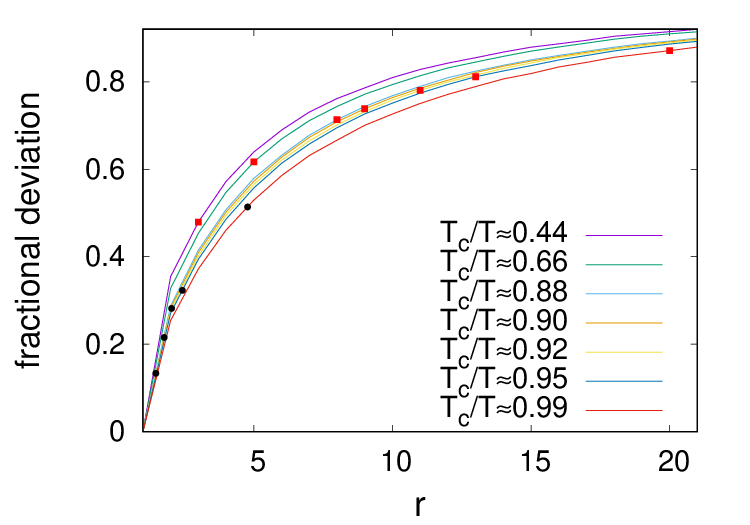}
    \caption{Geodesic rule deviation in $3D-$XY model with lattice distance}
   \label{fig8}
  \end{minipage}
\end{figure}

The Fig.\ref{fig7} and Fig.\ref{fig8} show the probability of deviation from the geodesic rule vs $r$  at various temperatures in two
and three dimensional systems respectively. The standard jacknife errors turned out to be within $1\%$ of the average values. We also carried out simulations by increasing the lattice size by $50\%$. In this case a very small increase in the deviation to the Geodesic rule was observed
for temperatures close to $T_c$. For simplicity this calculation was carried out along the $x$ and $y$ directions only. The results were found to be independent of directions. The results show that the probability of deviation increases with separation $r$ and temperature. The upper points on the curves for different temperatures correspond to the deviation at $r=\xi_{KM}$. Note that the deviation at $r=\xi_{KM}$ increases as we approach $T_c$ from above. The larger the $\xi_{KM}$, there are more possibilities to trace different paths. For comparison the deviation at the separations $r=\xi$ is also shown in these figures, indicated by black dots on the different curves. Though, this deviation is smaller than that at $r=\xi_{KM}$ but significant. For slower cooling, i.e for larger $\tau_Q$, the freezes out temperature will be closer to $T_c$. As a consequence the effect of Geodesic rule deviation will be larger. In the following, we look at other aspects of the KM.

\subsection{ Distribution of vortices in two spatial dimensions}

\begin{figure}[!tbp]
	\centering
	\begin{minipage}[b]{0.45\textwidth}
		\includegraphics[width=\textwidth]{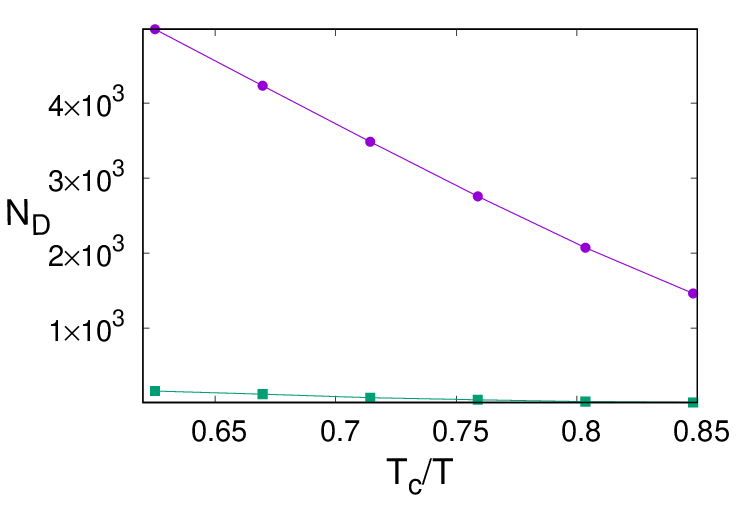}
		\caption{$N_D$ from Monte Carlo simulations for $2D-$XY model. The lower curve corresponds to estimate from KM }
		\label{2_dden}
	\end{minipage}
	\hspace{0.3cm}
	\begin{minipage}[b]{0.45\textwidth}
		\includegraphics[width=\textwidth]{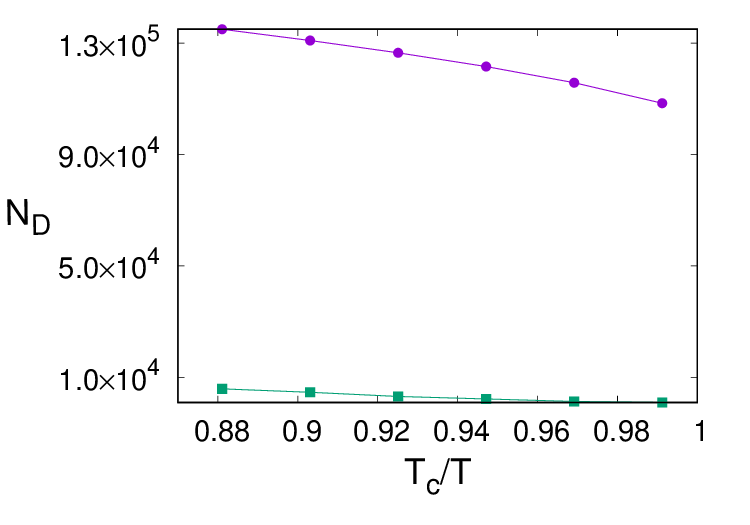}
		\caption{$N_D$ from Monte Carlo simulations for $3D-$XY model. The  lower curve corresponds to estimate from KM}
		\label{3_dden}
	\end{minipage}
\end{figure}

In two dimensions the defect density can be estimated using the KM, which turns out to be $(1/\sqrt{3}) \xi^{-2}_{KM}$\cite{Bowick:1992rz}. To test this
a sub-lattice of size $100\times 100$ is selected, small enough such that boundary condition does not affect the results. Inside the sub-lattice, each elementary squares (ES) is scanned for vortices. Given that the Geodesic rule is used to determine field variation between NN points, the net variation of $\theta$ along perimeter of an ES can have values $\delta\theta=-2\pi$, $0$ or $2\pi$. When $\delta\theta=-2\pi(2\pi)$ the ES is considered to have an anti-vortex(vortex) or none otherwise. Fig.\ref{2_dden} shows, $N_D$, the average of number of vortices and anti-vortices in the whole lattice. For comparison, each ES in three dimensional lattices are scanned for vortices. The results for $N_D$ are shown in Fig.\ref{3_dden}. In both cases, the number of vortices decreases with temperature. For larger lattices, no noticeable change in the defect density was observed. In Fig.\ref{2_dden} and Fig.\ref{3_dden}, KM predictions are also included, which underestimate $N_D$ by large margins for temperatures considered here. This discrepancy reduces when standard correlation length, $\xi$, and deviation of the Geodesic rule is taken into account. Even if deviation to the Geodesic rule is $50\%$, the KM estimates increase only by a factor of $2.5$. This suggests that at finite temperature, most of the defects are created via thermal fluctuations. Note that the defect
density will follow the above curves until the freeze out temperature. For lower temperatures, the density will remain
frozen until the system comes out of the non-adiabatic regime and the OP field starts to evolve again.

\subsection{Correlation between vortices and anti-vortices}

\begin{figure}[!tbp]
	\centering
	\begin{minipage}[b]{0.45\textwidth}
		\includegraphics[width=\textwidth]{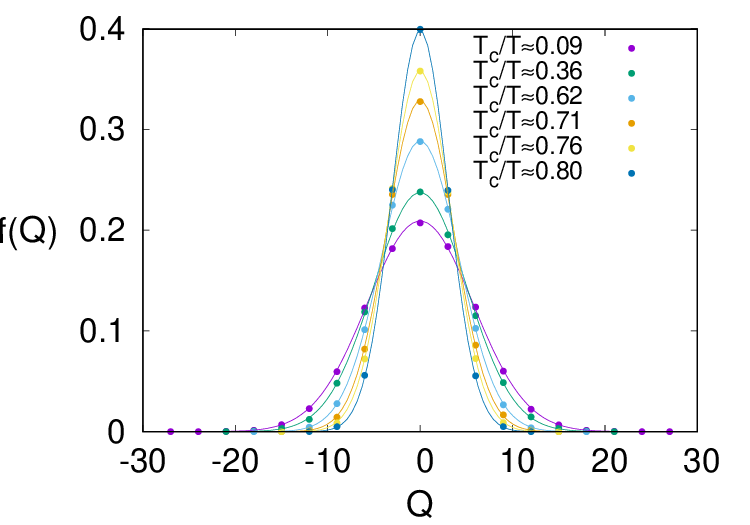}
		\caption{Distribution of the net winding number in the sublattice}
		\label{cdist}
	\end{minipage}
	\hspace{0.3cm}
	\begin{minipage}[b]{0.45\textwidth}
		\includegraphics[width=\textwidth]{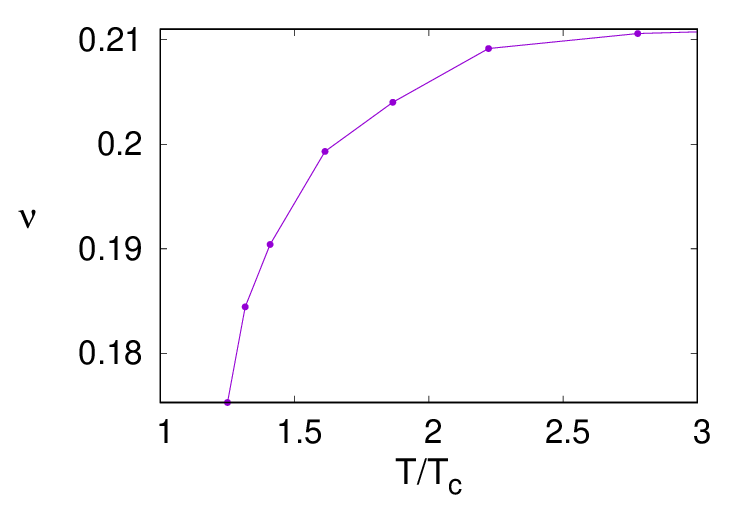}
		\caption{The exponent $\nu$}
		\label{nuex}
	\end{minipage}
\end{figure}

The Kibble Mechanism also predicts strong correlation among the vortices \cite{Digal:1998ak}. In a given area the standard deviation, $\sigma$, of difference in number of vortices and anti-vortices $Q=N_+ - N_-$ is given by $\sigma \sim N_D^{\nu}$, with $\nu=1/4$ \cite{Digal:1998ak,Vachaspati:1984dz}. On the other hand, for a completely random distribution of vortices $\nu=1/2$.
Fig.\ref{cdist} shows the distributions of net winding number in the sub-lattice  for various temperatures. The resulting values of $\nu$ are shown in Fig.\ref{nuex}. Note that a constant factor is to be taken into account while comparing these results with the Kibble Mechanism \cite{Digal:1998ak}. The resulting $\nu$ approaches the value $1/4$ when $T\to \infty$. This makes sense as the geodesic rule is imposed at the scale of lattice spacing and the NN spins are uncorrelated in this limit.  In the following we
study the Geodesic rule in $\phi^4$ theory.

\section{Geodesic rule in $\phi^4$ theory}
The lattice action for the $\phi^4$ theory with $U(1)$ symmetry is taken to be,
\begin{equation}
S_\phi = -\kappa\sum_{i,\mu}\left( \phi^\dag_i\phi_{i+\hat{\mu}} + h.c \right) + \sum_{i} \left[{1\over 2} \phi^\dag_i\phi_i + \lambda\left(\phi^\dag_i\phi_i-1\right)^2\right]
\label{S}
\end{equation} 
where $\mu=x,y,z$ and $\hat{\mu}$ is corresponding unit vector. The quartic coupling $\lambda$ is set to $\lambda=1$.
The NN coupling parameter $\kappa$ plays the role of inverse temperature in this case. The above model is similar to $O(2)-$spins except that the magnitude of the field can now vary. The partition function is given by
\begin{equation}
{\cal{Z}}= \int \prod_i d\phi_i^* d\phi_i~{\rm Exp}[-S_\phi].
\end{equation} 
In the thermodynamic limit the system undergoes a $2nd$ order phase transition at $\kappa_c$. The paramagnetic phase persists upto $\kappa_c$. For $\kappa > \kappa_c$ the system is found to be in the ferromagnetic phase. The Monte Carlo simulations in this case are
carried out using pseudo heat-bath algorithm. For details, see reference \cite{Bunk:1994xs}. The simulations are carried out for $60\times 60\times 60$ $3D$ lattice. 

\begin{figure}[h!]
	\centering
	{\rotatebox{360}{\includegraphics[width=0.48\hsize]
			{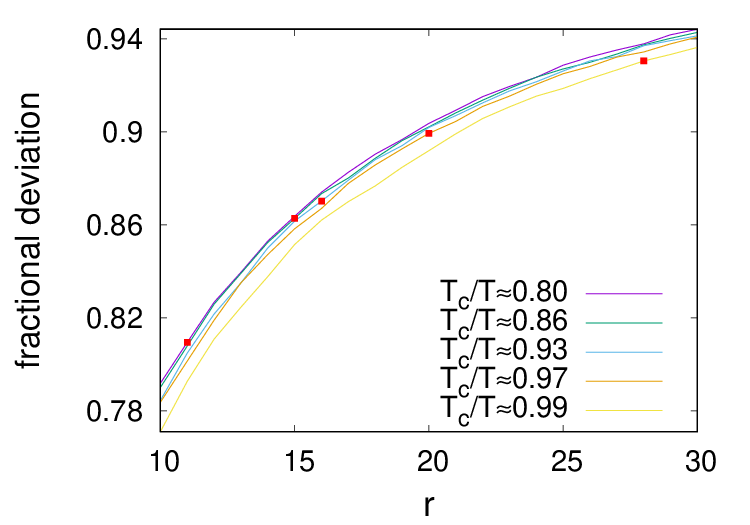}}
	}
	\caption{Geodesic rule deviation for $\phi^4$ theory}
	\label{fig11}
\end{figure}

At high temperatures, $\kappa < \kappa_c$, the volume and thermal average of $\phi$ is zero, however $\phi_i=0$ is suppressed due to vanishing measure. It is found that the image trajectory, of any field configuration over a linear stretch $r$, rarely passes through the origin ($\phi=0$) in the field space. Therefore, the phase $\theta$ of the field $\phi$ can be used for the computation of the probability of deviation from the Geodesic rule following the procedures adopted in the case of $O(2)-$spins. $\Delta\theta_{ij}$ and $\eta_{ij}$ are found as function of the separation $r=|{\bf r}_i-{\bf r}_j|$. As in the case of $O(2)$ spins, whenever $\Delta\theta_{ij}\ne \eta_{ij}$, it is considered to be deviation from the Geodesic rule. The Fig.\ref{fig11} shows the probability of deviation from the Geodesic rule as function of $r$ for different $\kappa$. The deviation increases with $r$ and temperature $1/\kappa$
similar to the case of $O(2)-$spins. The red squares on the curves correspond to the deviation at $r=\xi_{KM}$. 

\section{Conclusions}

We have studied the variations of field in the $XY-$model in two and three dimensions and $\phi^4$ theory in three dimensions using  Monte Carlo lattice simulations. For two dimensions the lattice size was taken to be $200\times 200$. For three dimensions we consider $60^3$ lattices. The computations are carried out above the critical temperature to check the validity of the Geodesic rule in the context of the Kibble-Zurek mechanism of defect formation. We compute the deviation to the geodesic rule considering different length scales such as $\xi_{KM}$ and $\xi$. The deviation is non-zero as long as the would be domain size is larger than the lattice spacing. At a given temperature, the deviation increases with length scale. Also as the temperature approaches $T_c$ from above we observe increase in the deviation.

We also studied the number of vortices and the net winding number in the case of $O(2)-$spins in two dimensions. The number of vortices is found to be significantly higher than what is 
expected from the Kibble-Zurek mechanism. The discrepancy reduces somewhat when the deviation from the geodesic
rule is incorporated. The results show that a large contribution comes from thermal production of vortices. This is supported by results of the distribution of the net winding number, which show that the standard deviation is much smaller than what is expected from the consideration of the Geodesic rule. This suggests that there is a significant pairing between the vortices and anti-vortices. It is expected that from topological considerations thermal fluctuations always create a pair of vortex-anti-vortex also at smaller separations. We mention here that in theories with gauge symmetries, there is no satisfactory argument for the Geodesic rule \cite{Rudaz:1992wy}. The trajectory of the field between two domains can be deformed by gauge transformations. The Monte Carlo simulation method presented here can be extended to settle the Geodesic rule in these theories.

\acknowledgements

We thank A. P. Balachandran, Ajit. M. Srivastava and Rajarshi Ray for valuable comments.

\end{document}